# Analogue of electromagnetically induced transparency with high Q factor in Metal-dielectric metamaterials

**BINGXIN HAN, ZHI HONG**

*Centre for THz Research, China Jiliang University, Hangzhou 310018, China*

**Abstract**: We investigated numerically the electromagnetically induced transparency (EIT) behavior with the maximum Q factor is up to 28000 in a metal-dielectric metamaterial. It composed of two layers, a metallic strip as bright mode stacked above a dielectric rod as another bright mode. The coherent coupling between the broad electric dipole induced by the metal strip and the narrow toroidal dipole induced by the dielectric rod leads to the destructive interference, then introduces a transparency window. Interestingly, the EIT window could be effectively influenced by varying parameters of the two layers and the thickness of middle spacer. This process can be explained by using two-particle model. Furthermore, the radiation loss is greatly suppressed within the EIT windows and the slow light effect is improved a lot. These excellent properties also provide an effective platform for the filters and sensitive chemical and biological sensors in the optical range.

1. Introduction

  Electromagnetically induced transparency (EIT) is an effect which originates from the interference of two different excitation pathways and then gives rise to a sharp peak within a broad transmission dip [1,2]. Over the past decade, plasmonic metamaterial (MM) [3-14] and dielectric metamaterial [15-22] based EIT-like phenomena have attracted enormous attention due to their brilliant performances in filters, slow light based devices and biosensors. EIT-like responses are usually achieved by two schemes: bright-dark mode coupling [3-9,15-22] and bright-bright mode coupling [10-16]. And for both two schemes, small wavelength detuning and large contrast of Q factor between the two coupling modes are necessary to achieve EIT response with high Q factor [23]. For metallic metamaterial systems, a variety of structures can be used to realize EIT, from cut wires [9], split-ring-resonators [12,13] to multi-layers structures [24]. However, apart from the inherent ohmic loss, the radiative loss is high as well in plasmonic MMs, which limit the Q factor of EIT response around 10. Comparing with metallic MM, all-dielectric MM possess the advantages of no ohmic loss and low possible radiative loss [17,18]. At present, EIT responses with high Q factors in dielectric MMs are all designed by using asymmetric structures and based on bright-dark mode coupling [17,18,20]. For example, Q value can be reached up to ~30000 in calculation and 483 in experiment in an all dielectric asymmetric rod-ring resonantors [17]. While all EIT resonators based on the bright-bright mode coupling have low Q factors [10-16] because of the difficulties of obtaining ultrahigh Q bright mode resonances and designing the bright-bright mode coupling structures. Thanks to the discovery of toroidal responses in MMs [25-32], ultrahigh Q resonance becomes possible [33-35]. On the other hand, based on the bright-dark mode coupling, stacked MMs have been designed for realizing the classical analog of EIT in plasmonic system [24,36-38]. In addition, based on bright-bright mode coupling, a stacked graphene-dielectric metamaterial also used to achieve tunable EIT effect [39]. These stacked structures provide an effective way for optical coupling of two resonances.

Herein, based on bright-bright mode coupling, we numerically investigated the realization of high Q EIT resonance in a stacked metal-dielectric MM factor in near-infrared regime. The stacked metallic-dielectric MM consists of two layers, one silver strip array supporting low Q electric dipole resonance which works as one bright mode, and another silicon rod array supporting ultra-high Q toroidal resonance, which works as another bright mode. The destructive interference between these two bright modes gives rise to the narrow bandwidth EIT response. To the best of our knowledge, this is the first time that the EIT-like response with ultra-high Q factor could be realized based on bright-bright mode coupling.

## 2. Design and structure

As shown in Fig. 1(a), the designed stacked metal-dielectric metamaterial consists of a metallic strip array and a dielectric silicon rod array, which are placed on top and bottom of the dielectric quartz spacer, respectively. The side view and top view of the unit cell for the metal-dielectric structure are shown in Fig. 1(b) and 1(c), respectively. The metallic strips made from silver, and its permittivity in the near-infrared is described by the Drude model with plasma frequency $\omega_{pl}=1.37\times10^{16}$ s$^{-1}$ and the damping constant $\omega_c=8.5\times10^{13}$ s$^{-1}$ [40]. Silicon is used as the material of dielectric resonator due to its high index and very small absorption. In our simulation, the refractive indices of silicon and the quartz spacer are set as $n$=3.45 and 1.46, respectively. Geometrical parameters of the unit cell in Fig. 1 are designed as follows: *l1* = 530 nm, *w1* = 150 nm, *h1* = 30 nm, *l2* = 830 nm, *w2* = 360nm, *h2* = 200nm, and the thickness of the quartz spacer is *t*. The electric field of the normal incident wave is polarized along x-axis. Both of the lattice periods in the x and y direction are 900 nm. Numerical simulations are conducted by using commercially available CST software at frequency-domain. In the simulations, periodic boundary conditions are considered both in the *x* and *y* directions, and perfectly matched layers(PMLs) are used in the wave propagating direction z.

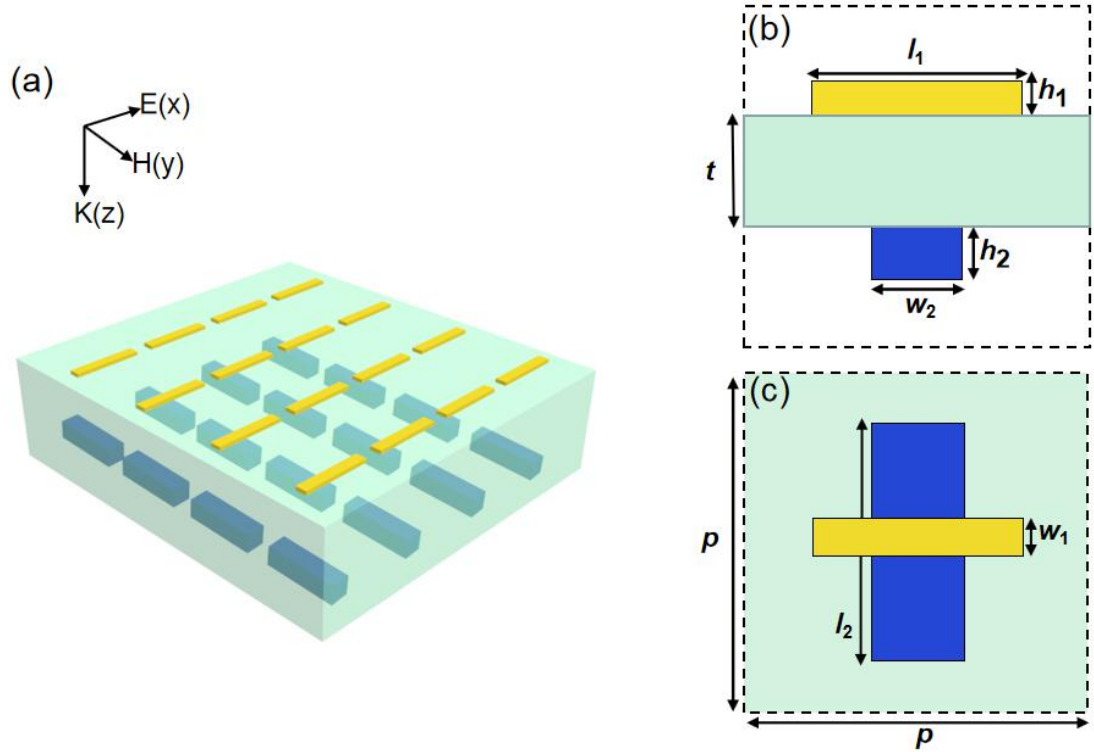

Fig.1 (a) Schematic of metal-dielectric metamaterial comprising periodical array of silver strips and silicon rods on the top and bottom of quartz spacer. (b) The side view and (c) top view of the unit cell.

Firstly, we investigated transmission characteristic of a single silver strips array with quartz substrate under the normal illumination of a x-polarized incident wave. The geometrical parameters of the silver strips are the same as in Fig. 1 ($l1$ = 530 nm, $w1$ = 150 nm, $h1$ = 30 nm) and a semi-infinite substrate is assumed in our simulation. As shown in Fig. 2(a), the transmission spectrum of the silver strips array shows a broad resonance with its dip at 1538 nm in calculated wavelength range from 1300 nm to 2000 nm, and the corresponding quality factor (Q) is 6. The surface current distribution at the broad resonant dip is displayed in Fig. 2(b). It exhibits a typical dipole mode characteristic with polarization charges flowing between the two ends of the resonator.

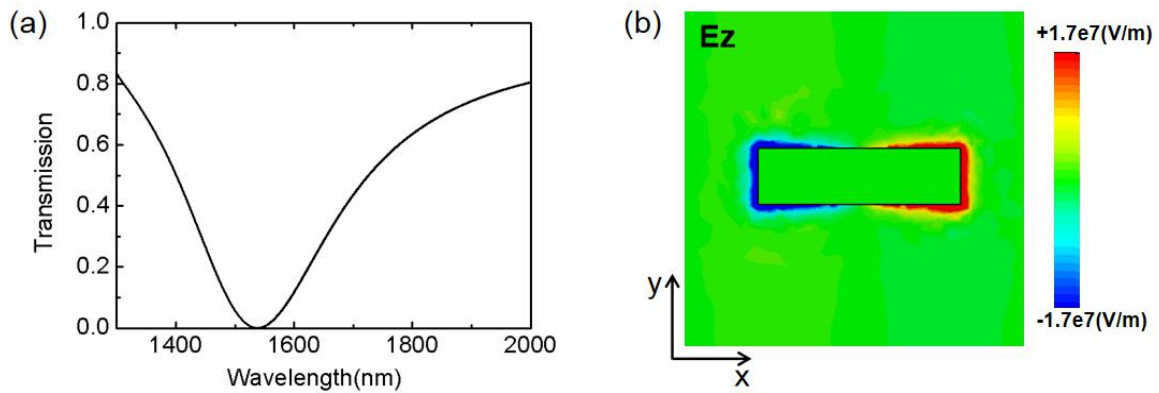

Fig. 2 (a) Transmission of a single silver strips array for the normal incidence by a plane wave with its polarization along the x-direction and (b) The z-component of the surface current distribution at the surface of the silver strip resonator at the wavelength of 1538 nm.

Next, we calculate the spectra of a single silicon rods array with quartz substrate under the same x-polarized incidence. The parameters of the silicon rods are as follows: $l2$ = 830 nm, $w2$ = 360 nm, $h2$ = 200 nm, respectively The simulated transmission in wavelength range from 1450 nm to 1650 nm is shown in Fig. 3(a), it can be noted that a sharp resonance with Q factor of 1405 appears at 1538 nm. The displacement currents in the x-y plane and magnetic field in the y-z plane at the resonance of 1538 nm are shown in Fig. 3(b) and (c), respectively. It can be seen that there are two opposite circular displacement currents on the upside and downside of the rod, each of them could produce a magnetic field, the corresponding z-component magnetic field of upside ring is along -z axis, and +z axis for downside ring, these induced magnetic field patterns generate a circular magnetic moment. This head-to-tail magnetic moment formed in the silicon rod indicates the excitation of toroidal dipole moment along the x-axis. In order to further access the role of the toroidal excitation in the observed resonance, decomposed scattered powers for multipole moments are calculated based on density of the induced current inside of the metamolecules by using a Cartesian coordinate system [25-27]. Figure 3(d) plots the five strongest scattering powers of multipoles around resonance wavelength of 1538 nm, which includes the electric dipole *Px*, magnetic dipole *My*, toroidal dipole *Tx*, electric quadrupole *Qe*, and magnetic quadrupole *Qm*, higher order dipoles can be ignored due to the extremely weak influence on the scattered intensity. The log scale in the y axis is chosen so as to reveal more clearly the contribution of the multipole terms. It is easy to see that in the vicinity of the resonance wavelength at 1538 nm, the toroidal dipole increases remarkably and dominates other multipoles in the far-field scattering power, which confirms the resonant mode as toroidal dipole [28-32]. Besides, the second largest dipole is magnetic quadrupole. This feature of contribution is similar to the contribution that many other high Q toroidal metamaterials has shown [29-30].

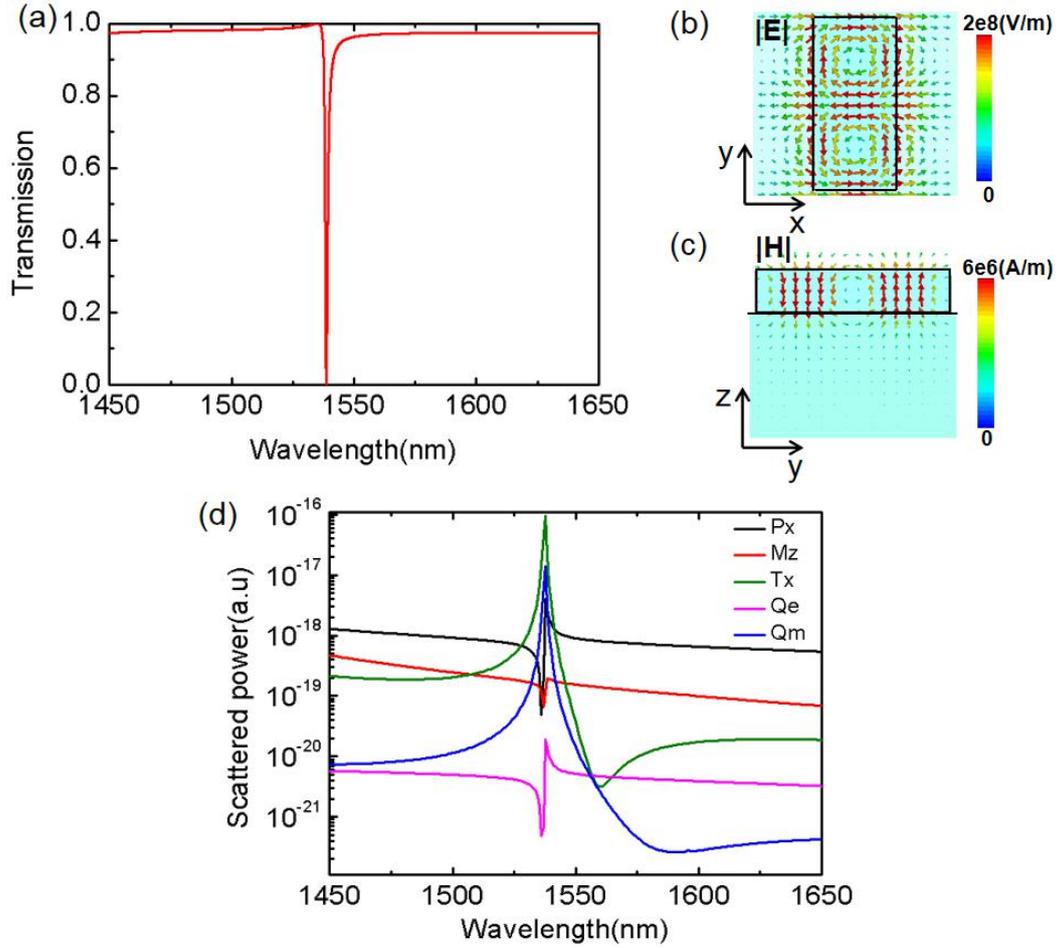

Fig. 3 (a) Transmission of a single silicon rods array under normal illumination by a x-polarized plane wave. (b), (c) Distributions of the electric and magnetic fields at λ = 1538 nm, respectively. (d) The full multipole decomposition of the first contributing five multipole moments of the metamaterial.

We further analyze the influences of geometrical parameters of the silicon rods, such as rod's length $l2$, width $w2$ and thickness $h2$, on the toroidal response, which is shown in Fig. 4. As the length of silicon rod $l2$ increases from 790 nm to 890 nm while the other parameters keep the same as that used in Fig. 3(a), a red shift of the toroidal response is observed in Fig. 4(a). At same time, Q value of the toroidal resonance increases greatly from xx to $1.7×10^4$. The multipole decomposition results (not shown here) of the scattered power indicated that the toroidal moment increases quickly as $l2$ increases, on the other hand, the near field distribution of electric and magnetic fields revealed that both the intra-toroidal dipole (generated inside the silicon rods) and inter-toroidal dipole (generated between the adjacent rods in the y direction) also increase, leading to the increase confinement of the electric and magnetic fields inside the rods and between the rods as well. Hence, the radiation into free space was greatly supressed. The maximum Q factor reaches up to $1.7×10^4$ when $l2$ = 880 nm.

Fig. 4(b) shows the results of the metamaterial calculated at different rod width $w2$. When the width varies from 320nm to 400nm while the rest parameters remain unchanged as well, the wavelength of the toroidal resonance also shows a red-shift, and the speed of the resonance wavelength red-shift is faster than the case of $l2$.

However, the Q factor does not change much. Quite similar results as the case of $w2$ can be obtained when the rod thickness $h2$ varies from 160 nm to 240 nm, the red-shifts of the resonance wavelength can be noticed and the Q factor almost remains the same, as shown in Fig. 4(c).

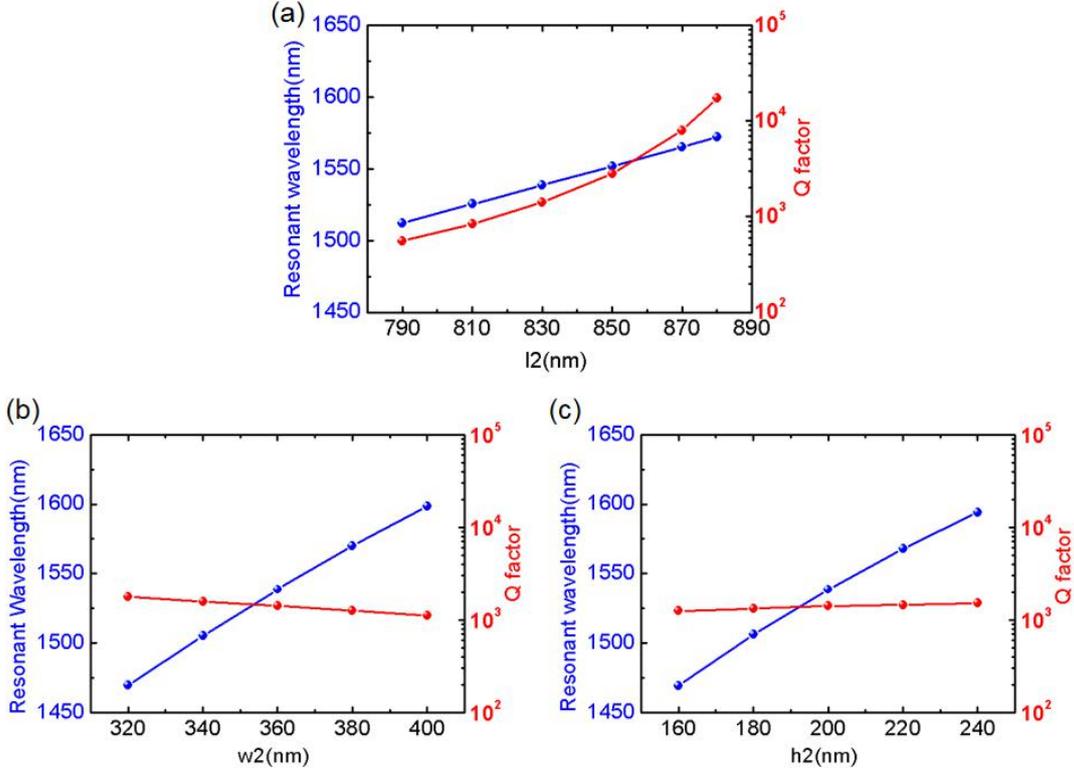

Fig. 4 The Q factor and resonance wavelength of toroidal dipole response with respect to the rod's parameters, (a) length $l2$ (when $w2$=360 nm and $h2$=200 nm), (b) width $w2$ (when $l2$=830 nm and $h2$=200 nm) and (c) thickness $h2$ (when $l2$=830 nm and $w2$=360 nm).

3. **EIT response based on bright-bright coupling and two-particle model**

From the analytic results above, it is clear that both of the single silver strips array and the silicon rods array are excited when illuminated by a normally incident light with its polarization along x-axis. In addition to the great contrast of Q factors between the two modes, the corresponding resonance wavelengths are almost overlapped, these conditions are necessary to achieve the sharp EIT response [23].

As the two layers are combined into the proposed metal-dielectric metamaterial, where the two layers are mutually perpendicular to each other as the shape of cross, an extremely sharp transparency window is observed in the calculated wavelength range from 1450nm to 1650nm, as shown in Fig. 5 (the blue solid line). The two dashed lines in Fig.5 are transmissions corresponding to the single metallic and dielectric arrays, respectively. Herein, the geometrical parameters are the same as in Fig.2(a) and 3(a) where $l2$ = 830nm, and the middle dielectric spacer $t$ is fixed as 900nm. The destructive interference between two bright modes, electric dipole and the toroidal dipole separately provided by silver strip and silicon rod leads to a transparent window. $P$ and $T$ are both along x-axis which makes the interference between them possible. The resonance wavelength of EIT response is at 1538nm and the corresponding Q factor is up to 3700. It is worth to mention that the Q factor of the EIT window can be greatly improved by increasing *l2* because of the larger contrast between the Q factors of the two bright modes.

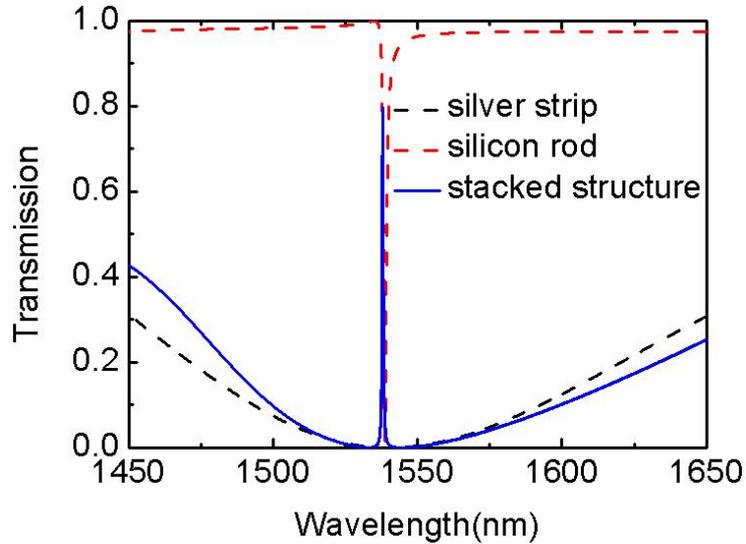

Fig. 5 Numerically simulated transmission spectra of a single silver strips array (dark dashed line), silicon rods array (red dashed line) and the metal-dielectric stacked metamaterial (blue solid line), respectively. The structure are illuminated by the x-polarized plane wave at normal incidence, the thickness of the dielectric spacer is 900 nm.

Furthermore, the influence of parameter *t* on the EIT resonance in the metal-dielectric metamaterial structure are investigated. The EIT response is changed when the middle layer with different thickness *t*, as shown in Fig.6. When the thickness of the middle spacer *t* increases from 300nm to 900nm, the transparency window which appears at the wavelength of 1538nm get narrower. As the result, the corresponding Q factor get improved from 100 to 3700 .

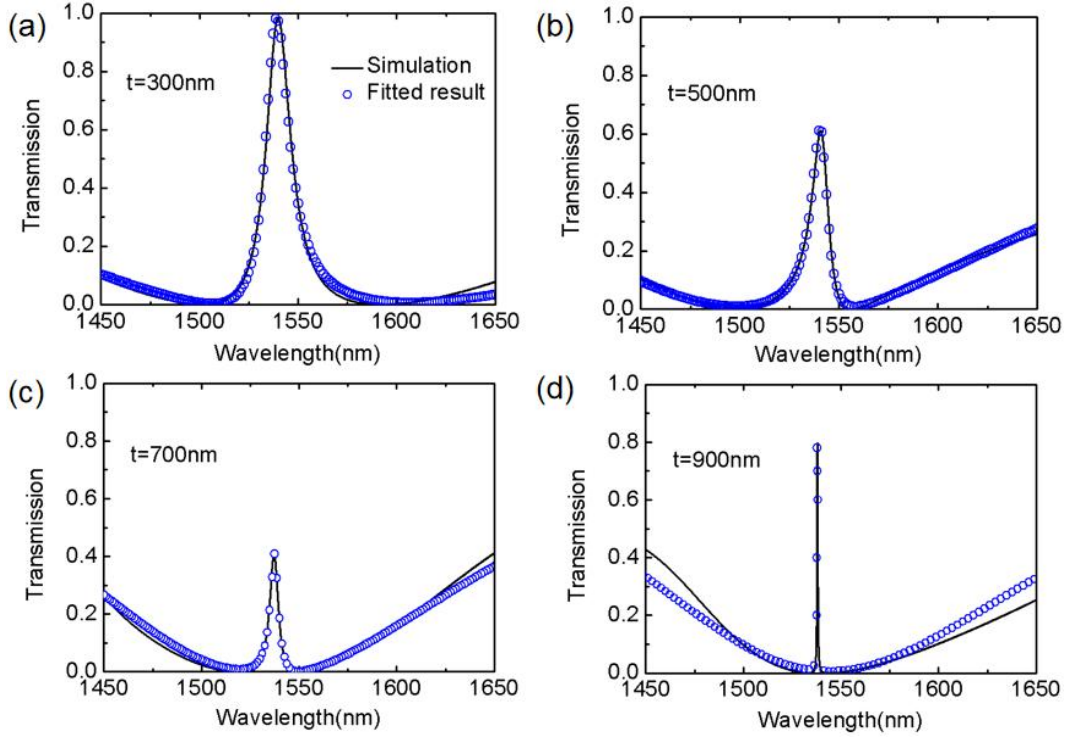

Fig. 6 The simulated (dark lines) and fitted (blue dotted lines) transmission spectra for different parameter *t*. (a) *t*=300nm,(b) *t*=500nm (c) *t*=700nm and (d) *t*=900nm, respectively.

To describe the impact of coherent coupling between two bright modes on the EIT in the metamaterial system, we introduce the two-particle model [11], where a particle represents the bright silver strip array, and a particle represents another bright silicon rod array. Both of particles can couple to the incident wave. In our case, the two particles have the same resonant frequency $\omega_1 = \omega_2$. These two particles satisfy the following set of equations:

$$\ddot{x}_1(t) + \gamma_1 \dot{x}_1(t) + \omega^2{}_1 x_1(t) + \kappa^2 x_2(t) = g_1 E / m_1 \qquad (1)$$

$$\ddot{x}_2(t) + \gamma_2 \dot{x}_2(t) + \omega^2{}_2 x_2(t) + \kappa^2 x_1(t) = g_2 E / m_2 \qquad (2)$$

where $x_1$; $x_2$, $\gamma_1$; $\gamma_2$, $\omega_1$; $\omega_2$, $m_1$; $m_2$ represent the displacements, damping rates, resonance frequencies and effective masses of two particles (the silver strip and silicon rod), respectively. $k$ is the coupling coefficient between two particles, while $g_1$ and $g_2$ are the geometric parameters indicating the coupling strength of two particles with incident wave. By solving the above coupled equations (1) and (2), we can obtain $x_1$ and $x_2$. Then, the effective electric susceptibility of the metamaterial can be written as

$$\chi_{eff} = \frac{P}{\varepsilon_0 E} = \frac{g_1 x_1 + g_2 x_2}{\varepsilon_0 E}$$
$$= \frac{K}{A^2 B}\left(\frac{A(B+1)\kappa^2 + A^2(\omega_2 - \omega_2^2) + B(\omega_2 - \omega_1^2) + i\omega(A^2\gamma_2 + B\gamma_1)}{\kappa^4 - (\omega^2 - \omega_1^2 + i\omega\gamma_1)(\omega_2 - \omega_2^2 + i\omega\gamma_2)}\right) \qquad (3)$$

Here, $K$ is the proportionality factor, and A = $g_1 / g_2$, B = $m_1 / m_2$. In order to validate the two-particle model, we fit 1-*Im* [$\chi_{eff}$] to the simulated transmission spectra as shown in Fig.6 (blue dotted lines). Obviously, the fitted results (blue dotted lines) show good agreements with the corresponding numerically simulated results. As expected, a decrease in the coupling coefficient $k$ from 45THz to 16.5THz and the Q factor of the transparency window increase from 100 to 6500 can be obtained with increasing $t$ from 300nm to 900nm, caused by the weaker coupling effect of two resonators. However, when $t$ is larger than 900nm, the coupling effect between two layers becomes much weaker because the distance is quite large, then the Q factor of the EIT resonance would not monotonically increased anymore. It is worth to mention that we do not show the case of $t$ is less than 300nm since the coupling coefficient $k$ is large and the Q factor of EIT window is low. There is small difference between calculated result and analytic result especially when $t$ is large, such as $t$=700nm and 900nm, that may be because the introduction of the F-P cavity effect.

Finally, we investigate the transmission properties of the proposed structure (here $t$ = 900nm) when the silver strip and silicon rod have certain displacement. As shown in Fig.6(a), the peak amplitudes and Q factors of the EIT window are calculated with the offset between two layers in x axis $s1$ changing from 0 to 450nm. Interestingly, we can find that the EIT window just has slight variations. It can be explained as that both of the electric dipole $Px$ and toroidal dipole $Tx$ have the same near-field strength during this x-axis shifting process. Fig.6(b) shows the calculated transmission peak and Q factors with the offset between two layers in y axis $s2$ changing from 0 to 450nm. It is obvious that the peak amplitudes of the EIT window decrease with the displacement changing from 0nm to 250nm. This tendency attributes to the position of the electric dipole $Px$ changes while the toroidal dipole $Tx$ remains. More specifically, the coupling strength between $Px$ and $Tx$ where the intra-coupling inside the silicon rod domains gets weaker when silver strip gets away from the central position. The EIT window disappears when $s1$=250nm. However, when the displacement shifts from 300nm to 450nm, the peak value of transmission gets increased again, but the Q factor becomes decreased from 21600 to 5180, which is because $Px$ starts to couple with $Tx$ where the inter-coupling between the silicon rods in adjacent period domains.

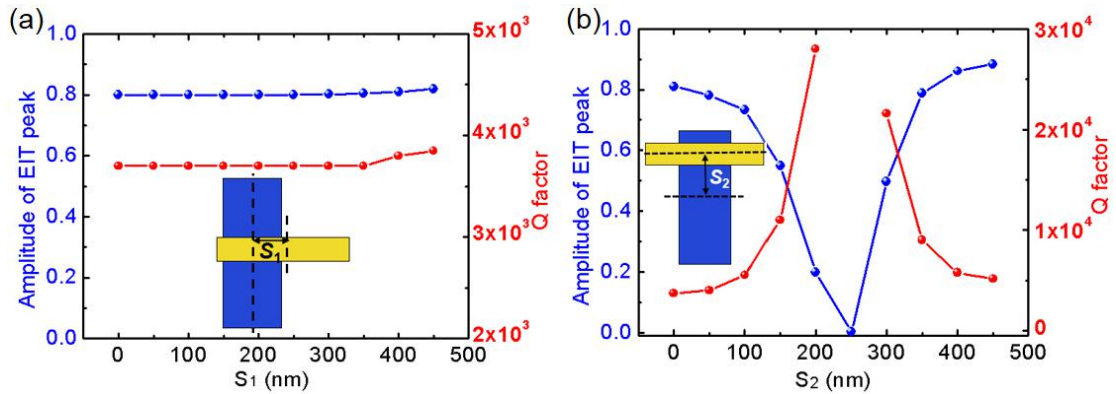

Fig.6 (a) The EIT peak and Q factors for silver trip array with different displacements $s1$ along x-axis and (b) $s2$ along y-axis in the proposed structure.

## 4. Conclusion

In summary, we have demonstrated the EIT response in a metal-dielectric metamaterial, and the unit cell of which consists a upper silver strip,a middle dielectric spacer and a lower silicon rod. The EIT resonance generated from the destructive interference between a broad electric dipole and a narrow toroidal resonance, both of them could be directly excited and served as two bright modes. These findings provide an useful access to realize EIT response with high Q factor. Moreover, the EIT window could be modulated by varying the thickness of the middle spacer and moving the silver strip along x and y axis. Our proposed structure provides an excellent platform for the biosensing and slow light.